# Towards a more sustainable academic publishing system

Mohsen Kayal[A], Jane Ballard[A], Ehsan Kayal[B]

[A]ENTROPIE, IRD, CNRS, IFREMER, Université de la Nouvelle-Calédonie, Université de la Réunion, Nouméa, New Caledonia

[B]Fédération de Recherche 2424 Sorbonne Université & Centre National pour la Recherche Scientifique, Station Biologique de Roscoff, Roscoff, France

**Abstract**

Communicating new scientific discoveries is key to human progress. Yet, this endeavor is hindered by monetary restrictions for publishing one's findings and accessing other scientists' reports. This process is further exacerbated by a large portion of publishing media owned by private, for-profit companies that do not reinject academic publishing benefits into the scientific community, in contrast with journals from scientific societies. As the academic world is not exempt from economic crises, new alternatives are necessary to support a fair publishing system for society. After summarizing the general issues of academic publishing today, we present several solutions at the levels of the individual scientist, the scientific community, and the publisher towards more sustainable scientific publishing. By providing a voice to the many scientists who are fundamental protagonists, yet often powerless witnesses, of the academic publishing system, and a roadmap for implementing solutions, this initiative can spark increased awareness and promote shifts towards impactful practices.

**Key Words**

Academic publishing, Scientific evaluation, Peer review, Publication fee, Open access, Copyright, Sustainability, Science for society.

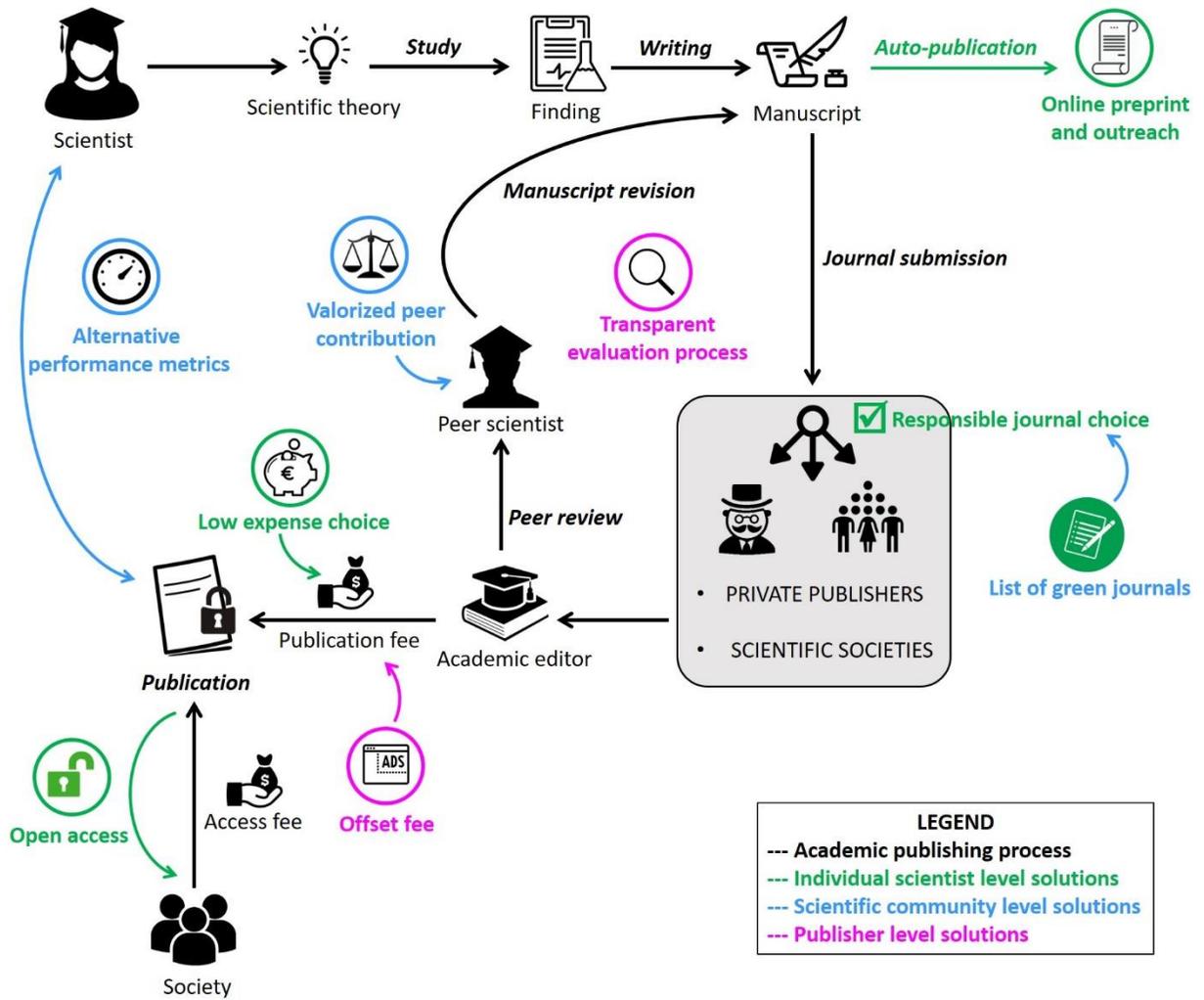

**Figure 1** The academic publishing process (black arrow path) and solutions for improvements at the individual scientist (green), scientific community (blue), and publisher (purple) levels.

**Today's academic publishing system**

Academic publishing is the predominant medium scientists use to describe and communicate their findings. It is also the main metric by which scientists are evaluated throughout their career, with frequency of publications and prestige of publishing media as a gauge of scientific achievements. In today's process of academic publishing, scientists provide, often even giving away author copyrights, valuable[1] scientific findings for free to journals, which then sell this knowledge to the public and other scientists (Liedes 1997, Odlyzko 1997, Bachrach 1998, Walker 1998, Smith 2006, Van Noorden 2013, Buranyi 2017). Indeed, most journals charge fees to both the authors for publication of, and to the readers for access to, scientific content, with some allowing higher publication fees to offset the access fee (a.k.a. open access; Smith 2006, Minet 2017, Zhang 2019) and vice versa. Academic publishers also rely heavily on the scientific community for the fastidious tasks[2] of manuscript selection and evaluation (i.e. editorship and peer review), tasks for which solicited scientists also receive no compensation[3] (Odlyzko 1997, Smith 2006, Aarssen and Lortie 2010, Van Noorden 2013, Schmitt 2014, Buranyi 2017, Zhang 2019). As such, any scientist (so-called "peer") can be solicited by a journal for evaluating articles submitted for publication by other scientists in the same field, with scientists contributing massively to editing and reviewing activities on a volunteer basis (Warne 2015, Riley and Jones 2016, Copiello 2018). This volunteer-based system of scientific publishing is not new, and was historically initiated by non-for-profit scientific societies with very small monetary power as a way to strengthen scientific endeavors. However, the academic publishing industry was quickly overtaken by for-profit companies which prospered on this highly profitable business model, thanks to both the decades of growing public investments in science and globalization of scientific literature (Odlyzko 1997, Walker 1998, Smith 2006, Van Noorden 2013, Schmitt 2014, Larivière et al. 2015, Buranyi 2017, Copiello 2018). This ongoing success-story of the publishing business contrasts profoundly with the dramatic decrease in funding and job opportunities experienced by the scientific community in the current era of austerity, precipitating many scientists into precariousness while being increasingly solicited for free contributions to an ever-expanding publishing industry (Aarssen and Lortie 2010, Schmitt 2014, McDonnell 2016, Herschberg et al. 2018, Ålund et al. 2020, RogueESR). Ironically, the rising competition among scientists resulting from funding scarcity has only exacerbated reliance on existing metrics of research excellency, predominantly based on the number of publications in, and editorship for, a handful of high-profile and often privately-owned, costly journals (Smith 2006, Aarssen and Lortie 2010, Heyman et al. 2016, Magistretti 2016, Buranyi 2017, Minet 2017), further draining scarce scientific resources into the cycle of publication and access fees. Here we provide a voice to the many scientists who are the fundamental protagonists yet often powerless witnesses of the academic publishing system, and identify several initiatives for stepping away from the current unethical model toward more sustainable academic publishing practices.

---

[1] Scientific findings rely on replicated and technical work which often necessitate time and costly resources.

[2] Scientific publications undergo meticulous evaluations by peer scientists acting as journal reviewers and editors, in charge of evaluating the adequacy of the design, execution, and interpretation of the study.

[3] The only benefits for such free-of-charge contributions happens when a scientist integrates into the editorial board of a high-profile journal, which can facilitate publishing in that journal and bring some notoriety among peers in the field (Walker 1998, Copiello 2018).

    Scientific activities rely heavily on funding allocated to individual scientists, research programs, and official institutions, in big part in response to publicly-funded calls for scientific investigation. Yet, paradoxically, dissemination of scientific findings predominantly lay in the hands of few for-profit publishing groups that largely benefit from the commercial exploitation of the knowledge produced by these publicly-funded endeavours (Smith 2006, Schmitt 2014,

Larivière et al. 2015, Buranyi 2017, Minet 2017, NewScientist 2018, Zhang 2019). The generalization of open-access publication of scientific findings funded by public research grants (Government of the Netherlands 2016, Minet 2017, Else 2018, OuvrirLaScience 2018), while a great initiative in itself, is in many ways amplifying scientific expenditure in publishing fees (Van Noorden 2013, Schmitt 2014), further wasting scarce public funds to lucrative journals. Neither ethical nor sustainable (Smith 2006), the conflictual scientific publishing system has generated intense debates within many academic institutions and scientific circles, with some recent strong initiatives withdrawing subscriptions to major publishing groups (Smith 2006, Larivière et al. 2015, Buranyi 2017, Minet 2017, Else 2018, Zhang 2019, The Cost of Knowledge), however with no result in a significant shift to a sustainable alternative.

**Towards a better publishing model**

When the system is drifting astray, individual and community choices can be a driving force towards a better pathway. Most scientists are already aware of the above-mentioned aberrations in the scientific publishing system. However, pushed by the inertia of the ongoing model and without a clear vision for alternatives, it is the path of least resistance to keep fueling conventional publishing practices as authors, reviewers and editors (Whitfield 2012, Heyman et al. 2016), despite this being a clear deviation from the fundamental commitments of the scientific communities for a better society (Odlyzko 1997, Bachrach et al. 1998, Walker 1998). Below we provide cues for an emancipation of the scientific community from the all-for-profit publishing system, and toward a circular and virtuous publishing model.

*At the level of individual scientists*, academics can avoid publishing in and evaluating for journals with unsustainable practices as prescribed by several previous initiatives, including The Cost of Knowledge initiated in 2012 (Whitfield 2012). A large spectrum of new alternatives has emerged, from community driven media with established ethical publishing strategies (e.g. Peer Community in, SCOAP3, Zenodo) to author-led, open peer-reviewing (Aarssen and Lortie 2010, F1000Research, Ideas in Ecology and Evolution, Peerage of Science) and public disobedience initiatives for free universal access to scientific literature (Sci-Hub), all of which could be more broadly embraced as a means to furthering scientific contributions to society. As a general rule, given the relatively low costs of archiving and distributing articles in electronic formats in the present digital era as printing and mailing are rapidly vanishing (Odlyzko 1997, Walker 1998), working with journals that charge minimal publication fees can encourage transition to a new economic model. Scientists can also communicate their engagement for a better academic publishing system by inquiring on equitable practices when invited to review articles or to stand on the editorial board by a journal. Stronger engagement demands some degree of compensation for each contribution (Warne 2015, Riley and Jones 2016, Copiello 2018), for example a discounted rate for publishing future articles as established by the PeerJ community (PeerJ) or, similar to practices found in other evaluations requiring scientific expertise (e.g. assessments of Ph.D. dissertations and research proposals), a monetary compensation as prescribed in ScienceMatters (ScienceMatters) and Ideas in Ecology and Evolution (Ideas in Ecology and Evolution). Such monetary compensations can be relinquished, donated to an external cause (e.g. via non-governmental organizations), or reinjected into scientific activities (research, hiring, etc.) at the will of the researcher. Compensation systems shifting away from monetization can consist of including reviewers with significant insights to the author list of articles (Aarssen and Lortie 2010). Making one's published articles available online (using preprint servers and personal or institutional web pages) and communicating scientific findings in other media that are directly accessible by the public (blog posts, photos and videos on various online platforms and so-called "social media" supported by peer scrutiny) can also help bypass the monopoly of academic publishers. Similarly, when looking for a paper locked under a paywall, contacting the authors directly rather than using the payment process will save precious research money while strengthening interactions among scientists. Fortunately, most, if not all authors gladly share their publications for free when contacted directly. Yet, despite these above mentioned alternatives, academic publishing, reviewing, and editing remain intrinsic tasks of scientists, and deciphering the publishing standards of each medium available on the market stands as an overwhelmingly challenging barrier for individual scientists. Therefore, scientist-level choices need to be backed up with community-level initiatives and engagement from the publishing industry.

*At the level of the scientific community*, creating an ethical label to orient authors on publisher practices, similar to the organic and fair-trade labels used in agriculture, can help scientists navigate through the various options and promote positive choices. Institutionalizing the practice of alternative methods, including funding and institutional requirements for open-access publications in 'fair'-labelled journals, and using research libraries to host open-access publications online (Walker 1998), will also move the needle. Sparse examples of such initiatives already exist, such as the U.S. government requiring research carried out by public employees to be published in the public domain without copyright transfer to a third party (Bachrach et al. 1998), and the European Commission launching an online platform to host publications resulting from its Horizon 2020 funding program as fully subsidized, open-access, and open peer-reviewed articles (Open Research Europe). Further broadening such practices would highly benefit scientific publishing. Similarly, the academic community can be the intermediary between individual scientists and academic publishers in defining a reference compensation system for editing and reviewing contributions (Warne 2015, Riley and Jones 2016, Copiello 2018). The broadening of such a compensation system shall facilitate transition to a self-supporting, fair, and economically circular publication system (Aarssen and Lortie 2010). In addition to reorienting part of the revenues to support scientific community contributors, the reviewer compensation system and the disclosure of reviewer identities and comments are promising incentives for promoting both better manuscript preparation by authors and more constructive evaluation by reviewers (Aarssen and Lortie 2010). Finally, finding new ways beyond journal prestige to assess scientific contributions and promote scientific rigor will help the community emancipate from the supremacy of established conventional journals (Odlyzko 1997, Aarssen and Lortie 2010, Whitfield 2012, Van Noorden 2013, Magistretti 2016, Else 2018). This includes an increased recognition of existing and emerging tools, such as using manuscript views and citation indices as metrics of article influence as prescribed by the PLOS initiative (Public Library of Science), and open community discussions and validations to measure article integrity as implemented by multiple publishing platforms (Aarssen and Lortie 2010, F1000Research, Peerage of Science, WikiLetters).

*At the publisher level*, endorsements of emerging initiatives and constructive alternatives through engagements dedicated to streamlining the publication process can support transparency and fairness. This includes a general paradigm shift in relation to copyrights, a central component of the conventional publishing industry (Bachrach et al. 1998, Walker 1998, Creative Commons). Generalizing the use of non-exclusive and royalty-free publisher licences, an already viable and ongoing system since 2003 (Public Library of Science), will allow authors to reproduce and distribute their findings independently, and encourage the publishers that actually provide additional contributions to the work performed by authors, such as improved content, complementary media design, or outreach support (Odlyzko 1997, Bachrach et al. 1998). Various journals already embrace open and author-led reviewing processes as ways to save the present demise of the scientific reviewing system (Aarssen and Lortie 1998, Peerage of Science), with several of them also publishing reviewer names and/or comments alongside research articles (Warne 2015, Eisen 2019, F1000Research, Ideas in Ecology and Evolution). This practice could further be expanded to publishing reviewers bio- and biblio-graphies with articles as ways to increase recognition of expert contributions (Riley and Jones 2016). Besides, alleviating publication fees can help make larger portions of funding available for research *per se*, which will constitute non-negligible support to the overall academic community. Recently, many journals with increasingly lower publication fees have emerged. In fact, scientific journals could publish high quality research for free if publication costs are covered by sponsoring and advertising, a system largely in use in diverse media, search engine, and mobile-app industries. From universities and laboratories to manufacturers and providers of scientific equipment, many academic institutions and research societies already use broad advertising, a portion of which can be displayed in specialized scientific publishing media. Advertisement-funded journals could constitute a win-win for profitable businesses that want to advertise products to potential and/or targeted clientele, while reducing burden on scientists in need of publishing at lower costs.

  ***Revolutionizing the academic publishing system*** inevitably involves several challenges in preserving publishing sustainability and scientific rigor. Many such barriers can be alleviated by changing incentives through steps taken at the individual, community, and publisher levels (Figure 1). In the present era of transition where old and new academic publishing models coexist, positive initiatives and institutionalized support can promote shifting to a new publishing system that is in line with the scientific commitments for accessible knowledge for the society as a whole. At the end of the road, academic publishers, scientific institutions, and individual scientists all need a durably operating publication system. Engaging collaborative initiatives among these protagonists is key to achieving sustainable publishing practices. Global responses to recent pandemics have shown that open science is possible and necessary (Wellcome 2016, 2020, International Coalition of Library Consortia 2020). Facing the social and environmental challenges of the 21st century (Ripple et al. 2017, Kayal et al. 2019), time has come to make such values a standard, rather than an exception, in the academic system.